\documentclass[aps,twocolumn,showpacs,superscriptaddress,floatfix]{revtex4}
\usepackage{graphicx}
\usepackage{psfrag}
\begin{document}
\newcommand{\beq}{\begin{equation}}
\newcommand{\eeq}{\end{equation}}
\newcommand{\ben}{\begin{eqnarray}}
\newcommand{\een}{\end{eqnarray}}
\newcommand{\bea}{\begin{array}}
\newcommand{\eea}{\end{array}}
\newcommand{\om}{(\omega )}
\newcommand{\bef}{\begin{figure}}
\newcommand{\eef}{\end{figure}}
\newcommand{\leg}[1]{\caption{\protect\rm{\protect\footnotesize{#1}}}}
\newcommand{\ew}[1]{\langle{#1}\rangle}
\newcommand{\be}[1]{\mid\!{#1}\!\mid}
\newcommand{\no}{\nonumber}
\newcommand{\etal}{{\em et~al }}
\newcommand{\geff}{g_{\mbox{\it{\scriptsize{eff}}}}}
\newcommand{\da}[1]{{#1}^\dagger}
\newcommand{\cf}{{\it cf.\/}\ }
\newcommand{\ie}{{\it i.e.\/}\ }

\title{Transition from isolated to overlapping resonances \\
       in the open system of interacting fermions}

\author{G.L.~Celardo}
\affiliation{Instituto de F\'{\i}sica, Universidad Aut\'{o}noma de
Puebla, Apartado Postal J-48, Puebla, Pue., 72570, M\'{e}xico}
\author{F.M.~Izrailev}
\affiliation{Instituto de F\'{\i}sica, Universidad Aut\'{o}noma de
Puebla, Apartado Postal J-48, Puebla, Pue., 72570, M\'{e}xico}\
\affiliation{ NSCL and Department of Physics and Astronomy, Michigan
State University, East Lansing, Michigan 48824-1321, USA. }
\author{V.G.~Zelevinsky}
\affiliation{ NSCL and Department of Physics and Astronomy, Michigan
State University, East Lansing, Michigan 48824-1321, USA. }
\author{G.P.~Berman}
\affiliation{ Theoretical Division and CNLS, Los Alamos National
Laboratory, Los Alamos, New Mexico 87545, USA. }

\begin{abstract}

We study the statistical properties of resonance widths and spacings
in an open system of interacting fermions. At the transition between
isolated and overlapping resonances, a radical change in the width
distribution occurs with segregation of broad (``super-radiant") and
narrow (``trapped") states. Our main interest is to reveal how this
transition is influenced by the onset of chaos in the internal
dynamics regulated by the strength of random two-body interaction.
In the transitional region, the width distribution and its variance,
as well as the distribution of spacings between resonances are
strongly affected by internal chaos. The results may be applied to
the analysis of neutron cross sections, as well as in the physics of
mesoscopic devices with strongly interacting electrons.

\end{abstract}

\date{\today}

\maketitle

Physics of marginally stable mesoscopic systems is an important part
of current scientific interest, both in nuclear physics and in
condensed matter physics. The center of attention in nuclear physics
moved to nuclei far from the valley of stability. Exotic nuclei have
low binding energy so that the continuum states are easily excited
by a weak external perturbation and appear as the closest virtual
states in quantum-mechanical calculations. The correct description
of interplay between reactions and intrinsic structure is also
important for practical applications connected to neutron cross
sections at low and intermediate energy. Mesoscopic solid-state
devices, such as quantum dots and quantum wires, are open being
explicitly coupled to the outside world. The problem of mesoscopic
conductance fluctuations \cite{beenakker97} as a part of more
general physics of quantum transport in many aspects is similar to
that of fluctuations in nuclear reactions \cite{weidenmueller90}. In
all such cases, a transition from a consideration of a closed model
system to a realistic open system in its interaction with continuum
becomes necessary.

A working instrument in the unified description of intrinsic states
and reactions, or decay channels, is an {\sl effective non-Hermitian
Hamiltonian} ${\cal H}$ derived by elimination of channel variables,
for example with the help of the projection method
\cite{feshbach5862,MW,rotter91,SZAP92,VZCSM05}. This Hamiltonian
describes the intrinsic dynamics of an open system; its eigenvalues
are {\sl complex energies}, $ {\cal E}_{j}=E_{j}-\frac{i}{2}\,
\Gamma_{j}$, where the width $\Gamma$ determines the lifetime of a
resonance, $\tau\sim\hbar/\Gamma$. The form of the imaginary part of
${\cal H}$, factorized in the amplitudes of entrance and exit
channels, is dictated by the unitarity of the scattering matrix. The
factorization brings in remarkable consequences observed in
numerical simulations \cite{kleinwachter85} and explained
theoretically in Refs.~\cite{Zel1}, where the analogy to optical
{\sl superradiance} \cite{dicke54} of coherently radiating atomic
emitters was pointed out.

When typical widths $\Gamma$ are small compared to the level
spacing, $D$, on the real energy axis, $\Gamma/D\ll 1$, the cross
sections display {\sl isolated} resonances. In the critical region
$\Gamma/D \approx 1$ with crossover to {\sl overlapping} resonances,
the width distribution displays sharp segregation of broad
short-lived ({\sl superradiant}) states and very narrow long-lived
({\sl trapped}) states \cite{SZAP92,Zel1,ISS94}. Correspondingly,
the distribution of poles of the scattering matrix undergoes a sharp
transition from one to two ``clouds" of poles in the complex plane
of resonance energies \cite{haake}. This crossover phenomenon is
universal. However, it is not yet known how the specific features of
this universal effect, and therefore the characteristic time scales,
depend on the internal interactions.

The region of overlapping resonances was and remains of special
interest starting from the seminal papers by Ericson
\cite{ericson63} where the general approach to fluctuations of cross
sections has been developed. In nuclear physics, this region until
now was not a subject of detailed study, partly because of lack of a
statistically reliable experimental body of data, however, it is
accessible in nuclear reactions for scattering of slow neutrons on
heavy nuclei. In solid states physics, strongly overlapping
resonances occur in the regime of the so-called {\it perfect
coupling}, where famous universal conductance fluctuations arise
\cite{beenakker97}. The similarity and differences between
conductance fluctuations and Ericson fluctuations in nuclear
reactions have been discussed in Ref. \cite{weidenmueller90}.
Recently, experiments in the region of overlapping resonances have
been performed in microwave cavities, see Ref.\cite{stock,richter}.

Below we study the statistical properties of resonance widths and
spacings in the transitional regime explicitly taking into account
the regular or chaotic dynamics inside the system \cite{ann,FI97}.
The local statistical properties of fully chaotic quantum states can
be identified with those of the Gaussian Orthogonal Ensemble (GOE)
\cite{brody}. Here, for the first time, the two-body random
intrinsic interaction is considered that allows one to vary the
degree of internal chaos in an open system. With the non-Hermitian
approach, random two-body dynamics was used in Ref.~\cite{Agassi} in
the limit of a very strong interaction, when the influence of the
regular mean field was neglected.

We consider a set of a large number, $N$, of intrinsic many-body
states $|i\rangle$ with the same exact quantum numbers. The states
are unstable being coupled to $M$ open decay channels. The dynamics
of the whole system is governed by an effective non-Hermitian
Hamiltonian \cite{MW,Zel1} given by a sum of two $N\times N$
matrices,
\begin{equation}
{\cal H}= H - \frac{i}{2}\, W\,;\,\,\,\,\,\,\,\,\,\,\, W_{ij}=\sum
_{c=1}^M A_{i}^{c}A_j^c;                         \label{1}
\end{equation}
the intrinsic states are labeled as $i,j,...$ and decay channels as
$a,b,c...$. Here $H$ describes Hermitian internal dynamics, while
$W$ contains the amplitudes $A^c_i$ coupling intrinsic states
$|i\rangle$ to the open channels $c$. Under time-reversal
invariance, both $H$ and $W$ are real symmetric matrices.

The Hermitian part, $H=H_0+V$, of the full Hamiltonian (\ref{1}) is
modeled by the so-called two-body random interaction (TBRI) of $n$
fermions distributed over $m$ single-particle states; the total
number of many-body states is $N=m!/[n!(m-n)!]$; in our simulations
$n=6,\;m=12,\;N=924$. In this model $H_0$ describes the mean field
part, where single-particle energies, $\epsilon_s$, are assumed to
have a Poissonian distribution of spacings, with the mean level
density $1/d_0$. The two-body interaction $V$ between the particles
\cite{FI97} is fixed by the variance of the {\sl two-body} random
matrix elements, $\langle V_{s_1,s_2;s_3,s_4}^2\rangle=v_0^2$. While
at $v_0=0$ we have a Poissonian spacing distribution $P(s)$ of
many-body states, for $d_0=0$ (infinitely strong interaction,
$v_0/d_{0} \rightarrow \infty$), $P(s)$ is close to the Wigner-Dyson
(WD) distribution typical for a chaotic system. The critical
interaction for the onset of strong chaos is given \cite{FI97} by
$v_{cr}/d_0 \approx 2(m-n)/N_s$, where
$N_s=n(m-n)[1+(n-1)(m-n-1)/4]$ is the number of directly coupled
many-body states in any row of the matrix $H_{ij}$. Thus, we have in
our model $v_{cr}/d_{0} \approx 1/20$, and often we have used the
value $v_0/d_0=1/30$ slightly less than $v_{cr}$. In parallel we
also consider the limiting case of $H$ as a member of the GOE
\cite{Zel1,VWZ} that corresponds to a {\sl many-body} interaction,
when the matrix elements are Gaussian random variables, $\langle
H_{ij}^2\rangle =1/N$ for $i \ne j$ and $\langle
H_{ij}^2\rangle=2/N$ for $i= j$. In order to compare with the GOE
case, when changing the strength $v_0$ of interaction, we
renormalize the model parameters to have the same mean level spacing
between many-body states at the center of the resulting energy
spectrum.

The real amplitudes $A_i^c$ are assumed to be random independent
Gaussian variables with zero mean and correlator $\langle A_i^c
A^{c'}_j\rangle=\delta_{ij} \delta^{cc'} \gamma^{c}/N$. This is
compatible with the GOE or TBRI model assumed for intrinsic dynamics
when generic many-body states coupled to continuum have complicated
structure, while the decay probes specific simple components of
these states related to few open channels. Even in the case of weak
intrinsic interaction, we need to have in mind that the states
$|i\rangle$ have certain values of exact constants of motion, such
as angular momentum and isospin in the nuclear case. At sufficiently
large dimension $N$, these states acquire {\sl geometric chaoticity}
\cite{ZV04} due to the almost random coupling of individual spins.
Therefore the ensemble of decay amplitudes is reasonable. The
parameters $\gamma^{c}/N$ with dimension of energy are unperturbed
partial widths which characterize the coupling to the channel $c$.
The normalization of this correlator is convenient if the energy
interval $ND$ covered by decaying states is finite. We neglect a
possible explicit energy dependence of amplitudes (the transitional
region is far from thresholds).

The effective Hamiltonian (\ref{1}) determines the reaction cross
sections, $\sigma^{ba}(E)\propto |S^{ba}(E)|^2$, or the scattering
matrix $S=\delta^{ba} -i T^{ba}$. The reaction amplitudes,
neglecting the smooth potential phases irrelevant for our purposes,
are expressed in terms of the amplitudes $A_{i}^{c}$:
\begin{equation}
 T^{ba}(E)=\sum_{i,j}^N A_i^b\left(\frac{1}{E-{\cal H}}
\right)_{ij} A_j^a.                  \label{2}
\end{equation}
The complex eigenvalues ${\cal E}_{j}$ of ${\cal H}$ coincide with
the poles of the $S$-matrix and, for small $\gamma^c$, determine
energies and widths of separated resonances. In the simulations we
consider an energy interval at the center of the spectrum of ${\cal
H}$ with the constant many-body level density $\rho(0)=D^{-1}$. As
$\gamma^c$ grows, the resonances start to overlap, the effective
parameter being $\kappa^{c}= \pi\gamma^{c}/2ND$. The transmission
coefficient in the channel $c$,
\begin{equation}
T^{c}=1-|\langle S^{cc}\rangle|^2=
\frac{4\kappa^{c}}{(1+\kappa^{c})^{2}},               \label{3}
\end{equation}
is maximal (equal to 1) at the critical point, $\kappa^{c}=1$, that
marks the transition to superradiance and trapping. We study the
statistical properties of resonance widths and spacings as a
function of the interaction between particles (ratio $v_0/d_0$) and
the continuum coupling parameter $\kappa$. For simplicity, we assume
$M$ equiprobable channels, $\kappa^{c}=\kappa$; the maximum value of
$M$ we considered was $M=10$. For each value of $\kappa$ we have
used $N_r=100$ realizations of the Hamiltonian matrices, with
further averaging over energy.

\begin{figure}[h!]
\vspace{-0.5cm}
\includegraphics[width=7.0cm,angle=-90]{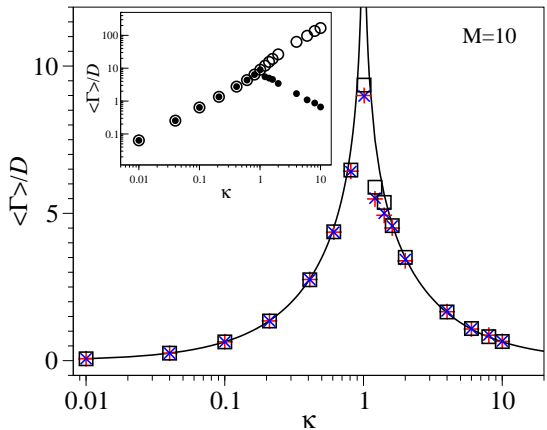}
\vspace{-0.1cm} \caption{(Color online) Average width versus the
coupling strength $\kappa$ for $M=10$. Solid curves show the
expression (\ref{5}), pluses refer to $v_0=0$, crosses to
$v_0=d_0/30$, squares to $v_0/d_0= \infty$. In all cases for
$\kappa>1$ the average was made over $N-M$ narrow resonances only.
More details are shown for $v_0=d_0/30$ in the inset, where open
circles correspond to the average over {\it all} $N$ resonances, and
full circles stand for the average over all resonances for
$\kappa<1$, and over $N-M$ narrow resonances for $\kappa>1$.}
\label{924GD10}
\end{figure}

At a critical value, $\kappa \approx 1 $, a segregation of the
resonance widths occurs \cite{kleinwachter85,Zel1,haake}, see inset
in Fig. \ref{924GD10}. The widths of $M$ resonances are increasing
at the expense of the remaining $N-M$ resonances. For weak coupling,
$\kappa \ll 1 $, the widths are given by diagonal matrix elements,
$\Gamma_{i}= \langle i|W |i \rangle= \sum_{c=1}^M (A_i^c)^2$, and
the mean width is $\langle\Gamma\rangle= \gamma M/N$. In the limit
of strong coupling, $\kappa \gg 1 $, the widths of $M$ broad
resonances converge to the non-zero eigenvalues of the matrix $W$
that has a rank $M$ due to its factorized structure which is
dictated by unitarity of the scattering matrix. As for the remaining
``trapped" $(N-M)$ states, their widths decrease $\propto 1/\gamma$.


In Fig.~\ref{924GD10} we show how the {\sl average width}
(normalized to the mean level spacing $D$) depends on $\kappa$ for
different values of $v_0/d_0$. The inset demonstrates a sharp change
in the width distribution at the transition point $\kappa =1$. The
results have been compared with the Moldauer-Simonius (MS) relation
\cite{M67,S74} (that follows also from our statistical assumptions)
\begin{equation}
\sum_{c=1}^M \ln(1-T^c)=-2 \pi \frac{\langle\Gamma \rangle }{D}.
                                           \label{4}
\end{equation}
For our case of $M$ equivalent channels,
\begin{equation}
\frac{\langle\Gamma\rangle}{D}= \frac {M}{\pi} \ln
\left|\frac{1+\kappa}{1-\kappa}\right|.              \label{5}
\end{equation}
Here we put the absolute value of the ratio under the logarithm in
order to extrapolate this expression beyond the transition point,
for $\kappa > 1$. The extrapolated MS-formula works well for any
interaction strength. For one channel, this result was obtained in
Refs.~\cite{Zel1,ISS94}; here we see that the MS-expression is also
valid for a large number of channels, independently of the
interaction strength. This agrees with the well known fact that
average quantities are in fact functions of the transmission
coefficient that is directly related to $\kappa$. This statement is
obvious for the GOE intrinsic dynamics where the results have to be
expressed in terms of orthogonal invariants. In our case this
follows from the assumed statistics of the continuum amplitudes when
we make this averaging first, independently of the degree of
intrinsic chaoticity \cite{CIZB07}. According to \cite{SFT99}, the
divergence of $\langle\Gamma\rangle$ at $\kappa=1$ is due to the
(non-integrable) power-law behavior for large $\Gamma$, see below;
in the numerical simulation, see Fig.~\ref{924GD10}, there is no
divergence because of the finite number of resonances although the
trend is clearly seen.


\begin{figure}[h!]
\vspace{-0.5cm}
\includegraphics[width=6.5cm,angle=-90]{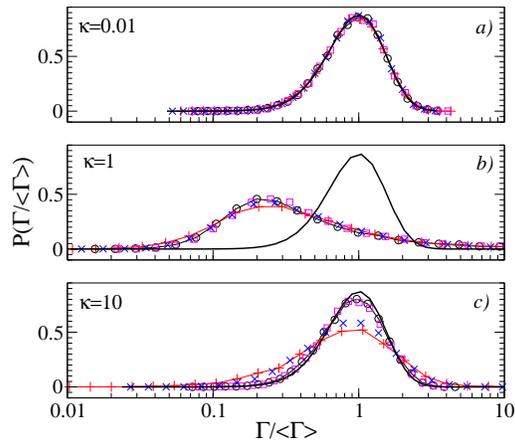}
\vspace{-0.1cm} \caption{(Color online) Width distribution for
$M=10$ and $\kappa=0.01;\,1;\,10$, with the same symbols as in
Fig.~\ref{924GD10}. Smooth curves are the $\chi^2_{10}$
distribution.} \label{924PGM10}
\end{figure}

The {\sl width distribution} $P(\Gamma)$ is shown in
Fig.~\ref{924PGM10} for different values of $v_0/d_0$ and
$\kappa=0.01,1,10$. At weak coupling, Fig.~2{\sl a}, the
conventional $\chi^2_{M}$-distribution is valid for any strength of
the interaction. However, when $\kappa$ increases, Fig.~2{\sl b} and
2{\sl c}, a clear dependence on the interaction strength emerges. As
first noted in Ref.~\cite{Mol68}, as $\kappa$ increases, $P(\Gamma)$
becomes  broader than the $\chi^2_{M}$ distribution. For large
$\kappa$, both for the GOE and for $v_0/d_0\rightarrow \infty$,
$P(\Gamma)$ is again given by the $\chi^2_M$ distribution, contrary
to the cases of the finite interaction strength; this reflects the
uniformity of properties of all fully chaotic intrinsic states.

Apart from the one-channel case studied in Ref.~\cite{mizutori93}, a
general analytical treatment of the width distribution is a
difficult task even for simplified random matrix models. In the case
of GOE for the Hermitian part $H$, the width distribution was found
in Ref.~\cite{SFT99}, neglecting the emergence of broad resonances
in the regime of strong overlap at $\kappa > 1$. Our results for
$v_0/d_0 = \infty$ correspond well to the analytical predictions. In
particular, the tails of $P(\Gamma)$ decrease $\sim \Gamma ^{-2}$
that leads to the divergence of $\langle\Gamma\rangle$ in agreement
with the MS-formula.

The (normalized) {\sl variance of the resonance widths}, see Fig.~3
for $M=10$ and $\kappa = 0.8$, strongly depends on the interaction
between particles. Our results confirm the analytical estimate for
the transition to strong chaos that can be associated with the
WD-distribution of level spacings. Note, however, that the
transition is quite smooth, and even for a relatively strong
interaction there is a deviation of the variance from the GOE value.
We would like to stress that the level spacing distribution, as the
weakest signature of quantum chaos \cite{ann}, turns out to be
rather insensitive to the transition. Indeed, the data in the inset
for $\kappa=0$ roughly correspond to the WD-distribution, although
once the continuum coupling is switched on, the variance of the
resonance widths is very different from that predicted by the GOE.

\begin{figure}[h!]
\vspace{-0.5cm}
\includegraphics[width=6cm,angle=-90]{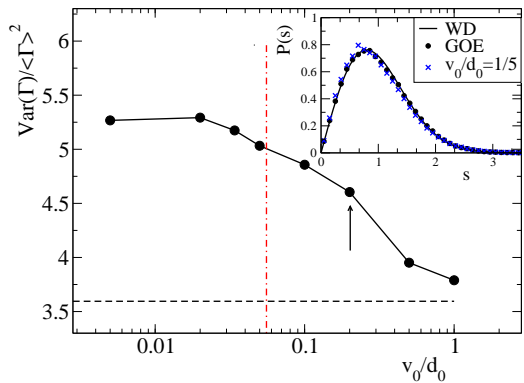}
\vspace{-0.2cm} \caption{(Color online) Dimensionless variance of
the widths versus $v_0/d_0$ for $M=10$ (connected circles): the GOE
value is shown by horizontal dashed line; the vertical dot-dashed
line marks the value $v_0=v_{cr}$ corresponding to the onset of
chaos. In the inset the level spacing distribution is shown for
$\kappa=0$ and $v_0/d_0=0.2$ (crosses), see the arrow in the main
part, and for the GOE (circles). The smooth curve is the
WD-distribution. } \label{PG12}
\end{figure}


\begin{figure}[h!]
\vspace{-0.5cm}
\includegraphics[width=5.0cm,height=8.8cm,angle=-90]{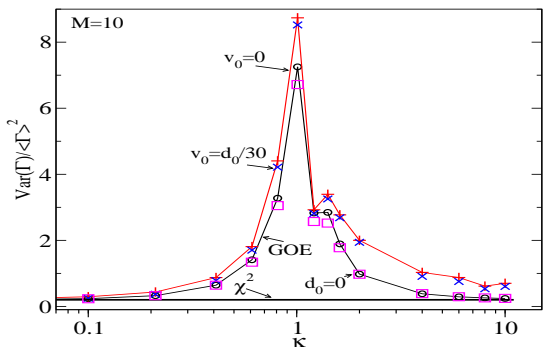}
\vspace{-0.1cm} \caption{(Color online) Dimensionless variance of
the width vs $\kappa$ for $M=10$. The dependence due to the
$\chi^2_M$ distribution is shown by the horizontal line; the symbols
are the same as in Fig.~\ref{924GD10}. Non-smooth dependence for
$\kappa > 1$ is due to a shift of the critical point for $M \gg 1$
from $\kappa=1$ to $\kappa=\kappa_c > 1$, see Ref. \cite{haake}.}
\label{924chiG}
\end{figure}

It is instructive to show how the variance of the widths depends on
the continuum coupling, Fig.~\ref{924chiG}. The prediction of the
$\chi^2_M$ distribution that the variance is $2/M$ independently of
$\kappa$ is correct for very weak and very strong coupling. As
expected, the variance takes its maximal value at the transition
point $\kappa = 1$ (it diverges for $N\rightarrow\infty$). As for
the dependence on $v_0$, the less chaotic is the intrinsic dynamics
and therefore greater a possible difference in the structure of
decaying states, the larger is the width variance. For the GOE and
$v_0/d_0\rightarrow \infty$, the system returns to the $\chi^2_M$
distribution, with an increase of $\kappa$. This is not true for a
finite value of $v_0/d_0$, in agreement with the results of
Ref.~\cite{Seligman1}, where justification of the broadening of the
width distribution in the Poisson case (equivalent to $v_0=0$) as
compared to the GOE was given for $M=1$.


\begin{figure}[h!]
\vspace{-0.5cm}
\includegraphics[width=7cm,angle=-90]{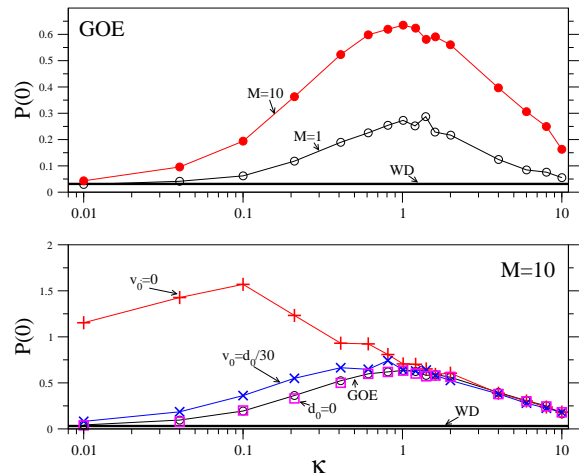}
\vspace{-0.1cm} \caption{(Color online) The probability $P(0)$ of
finding the level spacing $s<0.04$ (between the real parts of poles
of the scattering matrix, in units of mean level spacing) vs
$\kappa$. Upper panel: GOE for $M=1$ (open circles) and $M=10$ (full
circles); lower panel: $M=10$ with the same symbols as in
Fig.~\ref{924GD10}. The value due to the WD-distribution is given by
the horizontal line.} \label{924PSM}
\end{figure}

Finally, we analyze how the {\sl level repulsion} (in our case the
repulsion of the centroids of resonances along the real energy axis)
depends on the continuum coupling and on the intrinsic interaction.
The characteristic quantity is the probability of the level
proximity, $P(s\rightarrow 0)$. It is finite for the Poisson
distribution and vanishes linearly for the GOE. A weakening of the
level repulsion with an increase of the continuum coupling (or the
number of channels $M$) was noted as a generic property in
Ref.~\cite{Mol68}. The physical explanation (energy uncertainty of
quasistationary states) was given in \cite{Zel1}. Our detailed data,
Fig.~\ref{924PSM}, confirm these results and show the dependence on
$v_0/d_0$. For non-interacting particles, $v_0=0$, the results are
consistent with those of Ref.~\cite{Seligman2}, where it was argued
that, for the Poissonian intrinsic Hamiltonian $H$, the distribution
$P(s)$ becomes similar to that of a chaotic system due to the
perturbation induced by the coupling to the continuum. Our data
clearly show that with increase of intrinsic chaos the level
repulsion grows, however, for finite $v_0$ always remains weaker
than for the GOE case.

In conclusion, we have studied statistics of complex energies
(resonance widths and spacings) for a generic fermion system coupled
to open decay channels. For the first time we carefully followed
various signatures of the crossover from isolated to overlapping
resonances in dependence on the inter-particle interaction. We show
that the Moldauer-Simonius expression for average widths is valid
also in the region $\kappa > 1$, provided the $M$ widest resonances
are excluded. Also, we found that the average widths are insensitive
to the interaction strength. On the other hand, the width variance
depends on the degree of intrinsic chaos at the transition point
$\kappa \approx 1$ in a non-trivial way. Even if for a closed system
($\kappa=0$) the level spacing distribution is of the Wigner-Dyson
form, the width distribution for a perfect coupling to the continuum
($\kappa \approx 1$) is very different from the GOE predictions.

Our study is based on the two-body random nature of the
inter-particle interaction. However, since we explored the central
part of the energy band where the eigenstates are delocalized, one
can expect similar statistical properties of resonances if one
substitutes the intrinsic interaction $V$ by the GOE matrix, with a
proper rescaling of the interacting strength $v_0$. This may allow
for an accurate analytical analysis. Another extension should
consider realistic shell-model interactions which bring, at a high
level density, the results close to those for banded random
matrices. The results on statistical properties of the reaction
cross sections, including deviations from the theory of Eriscon
fluctuations, are reported elsewhere \cite{CIZB07}. The obtained
results can be applied to neutron resonances in nuclei and to open
mesoscopic systems in the crossover region, where we found strong
deviations from the conventional level statistics and $\chi^{2}$
width distribution. Although it might be difficult to extract from
the experiment the exact statistics of poles in the complex plane
(which would be extremely desirable), at least the trends of
distributions along the road from isolated to overlapping resonances
certainly can be studied in well measured neutron resonances on
heavy nuclei at energies further away from threshold. The
experiments with microwave billiards (analog of one-body
shape-dependent chaos) at variable continuum coupling can shed light
on some features of the process of width collectivization and
segregation. Similar effects are expected in mesoscopic conductance
fluctuations which will be discussed in our forthcoming publication.

We acknowledge useful discussions with T.~Gorin, T.~Kawano, A.
Richter, D. Savin, and V.~Sokolov. The work was supported by the NSF
grants PHY-0244453 and PHY-0555366. The work by G.P.B. was carried
out under the auspices of the National Nuclear Security
Administration of the U.S. Department of Energy at Los Alamos
National Laboratory under Contract No. DE-AC52-06NA25396. F.M.I.
acknowledges the support by the CONACYT (M\'exico) grant No~43730.

\end{document}